# Porous silicon carbide and aluminum oxide with unidirectional open porosity as model target materials for radioisotope beam production

M.Czapski[a], T.Stora[a], C.Tardivat[b], S.Deville[b], R.Santos Augusto[a], J.Leloup[b], F.Bouville[b], R.Fernandes Luis[c]

[a] CERN, Genève 23 CH-1211, Switzerland
[b] Lab. de Synthèse et Fonctionnalisation des Céramiques, CNRS/Saint-Gobain, Av. Jauffret 84306 Cavaillon, France
[c] Univ. Técnica de Lisboa Estrada Nacional 10, 2686-953 Sacavem, Loures, Portugal

**Abstract**

New silicon carbide (SiC) and aluminum oxide ($Al_2O_3$) of a tailor-made microstructure were produced using the ice-templating technique, which permits controlled pore formation conditions within the material. These prototypes will serve to verify aging of the new advanced target materials under irradiation with proton beams. Before this, the evaluation of their mechanical integrity was made based on the energy deposition spectra produced by FLUKA codes.

## 1. Introduction

At ISOL (Isotope Separation On-Line) facilities, a variety of radioactive ion beams (RIB) is produced by bombarding a thick solid or liquid target with highly energetic particles to initiate nuclear reactions (i.e. fission, fragmentation, spallation) in the material. One of such facilities is ISOLDE located at CERN [1], which has a long tradition of using different target materials, which are subjected to pulsed beams of 1.4 GeV protons from the Proton Synchrotron Booster (PSB). A search for more intense isotope beams with shorter half lives requires the use of materials characterized by faster release i.e. having better diffusion and effusion properties.

Among the different ISOLDE targets, there are molten metals, powders and foils, refractory carbides and oxides. The latter materials are particularly of interest due to their stability at high temperature (at which targets operate) in comparison to pure metals [2]. Two compounds here proposed can serve as model target materials among these groups i.e. silicon carbide (SiC) and aluminum oxide ($Al_2O_3$). Recent studies [3] showed that further modifications of their microstructure by decreasing the grain size from micron region to tens of nanometers and by controlling the porosity fraction, can alter significantly the release properties of these materials. Following these results, nowadays slip-casted SiC of 63% porosity are a part of ISOLDE regular target materials. On the other hand such changes can also influence the material stability, which can lead to a significant drop in target's lifetime. Therefore further investigations of evolution of their microstructure in irradiative environment should be done. In this study, a synthesis of different controlled sub-microstructures of the above selected model materials with the ice-templating method is presented. Furthermore, mechanical behavior is estimated using energy deposition spectra deduced from FLUKA codes to assess mechanical integrity under irradiation. This analysis will be used to verify further experimental results of irradiation with two proton beams of 1.4 GeV and 440 GeV.

## 2. Experimental

### 2.1. Sample preparation

The samples were prepared using the ice-templating technique [4–6], which allows one to produce materials with open directional porosity. In our case, initial slurries were made from SiC (Hexoloy® and Crystar®) and $Al_2O_3$(Ceralox®) and were frozen on the cold support with two different ice-front velocities. Temperature gradient within the slurry (trapped inside a Teflon mold) froze the water inside in direction perpendicular to the support's surface (Fig. 1).

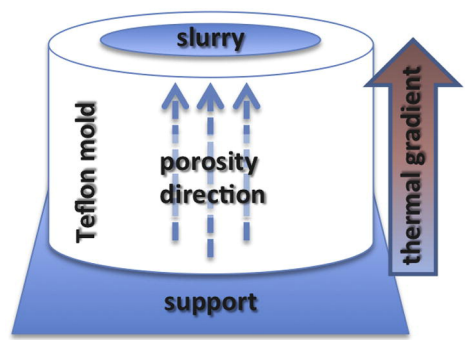

Fig. 1. Schematic of the experimental setup used to directionally freeze the slurries.

After the freezing process, the samples are put into a freeze-drier where, under lowered pressure, water crystals sublimate from the green body. In this way, the introduced porosity is of an open type. Two different freezing rates produce samples with two different pore sizes, without changing the porosity fraction. The later property can be controlled by changing the amount of powder in the initial slurry.

We were also able to modify pore shape with selected additives (NaOH, zirconium acetate (ZRA) complex). Table 1 summarizes the parameters and the properties they modify. Fig. 2 shows SEM images of different pore formations with a summary of structure modification. The so-prepared green bodies are ready for sintering. Porosity content achieved by this technique reaches from 25–90% [4]. Our prototypes have porosity content from 40–70% with pore sizes ranging between 1.1–26 μm ($Al_2O_3$), 11–27 μm (SiC Hexoloy) and 62–86 μm (SiC Crystar).

Table 1. List of parameters which directly control the final pore formation in a slurry.

| **Experimental parameter** | **Green compact property** |
| --- | --- |
| Powder fraction in slurry | Porosity fraction |
| Freezing speed | Pore size |
| Additives | Pore structure |

## 2.2. Mechanical calculations

In order to estimate the mechanical integrity upon irradiation, which allows the study of the impact of an irradiative environment on aging of the preselected prototypes, a simple mechanical analysis of the system was made. The irradiations took place at ISOLDE and HiRadMat [7] using CERN pulsed beams at 1.4 GeV (PSB) and at 440 GeV (Super Proton Synchrotron, SPS), respectively. The total number of protons used in both cases is equal to 1016 protons, due to HiRadMat facility limitations. Other parameters such as gaussian σ beam radii (3.5 mm and 2.0 mm for ISOLDE and HiRadMat, respectively) and repetition rates (1.2 s and 40 s, respectively) are predetermined by the accelerator operation (PSB and SPS, respectively).

FLUKA codes [8] were used to simulate irradiation of cylindrical samples of 1 cm in diameter and 4 cm long (which is an approximation of tightly stacked pellets of the real experiment) with ice-templated channels parallel to cylinder's long axis which is also the axis of the proton beam. Fig. 3 shows the energy deposition spectra in both materials simulating the conditions of ISOLDE and HiRadMat. For both irradiation types, energy is deposited radially in a similar manner with the maximum value located in the core of the sample. Assuming that the biggest thermal shock will occur in this region, the point of maximum deposited energy per volume per primary particles (ppp) ($E_{maxfluka}$) was used for further calculation. The energy deposited in the material per volume ($E_{dep}$) can be calculated from the formula:

$$E_{dep} = E_{maxfluka} \cdot N_{ppp} \quad (1)$$

where $N_{ppp}$ is number of protons per pulse ($3 \cdot 10^{13}$ and $8 \cdot 10^{12}$ for ISOLDE and HiRadMat respectively). Under this assumption and taking different repetition rates (t), we can evaluate the power deposited (P) in the material volume (V) as:

$$P = E_{dept} \cdot V. \quad (2)$$

The temperature increase per pulse ($\Delta T$) can be then calculated from the formula:

$$\Delta T = \frac{E_{dep}}{\rho^* C_p} \quad (3)$$

where $C_p$ is the specific heat and $\rho*$ is the reduced density calculated as:

$$\rho^* = \rho \cdot (1 - \text{porosity}), \quad (4)$$

where $\rho$ is the nominal density of the material. In this study we show calculations for the porosity fraction of 50% in comparison to unmodified material. By knowing the latter, thermal stresses $\sigma$ can be evaluated from:

$$\sigma = \Delta T \cdot E^*_{young} \cdot \alpha \quad (5)$$

where $E^*_{young}$ is the reduced Young modulus and $\alpha$ is the thermal expansion coefficient. It is assumed that the cylinder is subjected to deformation along its long axis. In this scenario the walls undergo only axial extension and compression [9]. The reduced Young modulus can be calculated as:

$$E^*_{young} = E_{young} \; \frac{\rho^*}{\rho} \quad (6)$$

where $E_{young}$ refers to the Young modulus of the nominal density material. Generated stresses should be lower than the compression strength of the material, therefore the material should remain intact. If we assume that the properties of ice-templated ceramics are similar to ceramic foams we can calculate the (reduced) strength in compression ($\sigma^*_s$) which is related to the cell wall strength ($\sigma_s$) [10]:

$$\sigma^*_s = 0.2 \; \sigma_s \left(\frac{\rho^*}{\rho}\right)^{\frac{3}{2}} \quad (7)$$

Table 2 summarizes all the thermo-mechanical properties of materials used in this study. The thermo-mechanical analysis results are presented in Table 3. The data reported in the last two columns show that in average the thermal shock is much below the reduced strength of the ice-templated structure, which suggests that its mechanical integrity should be guaranteed.

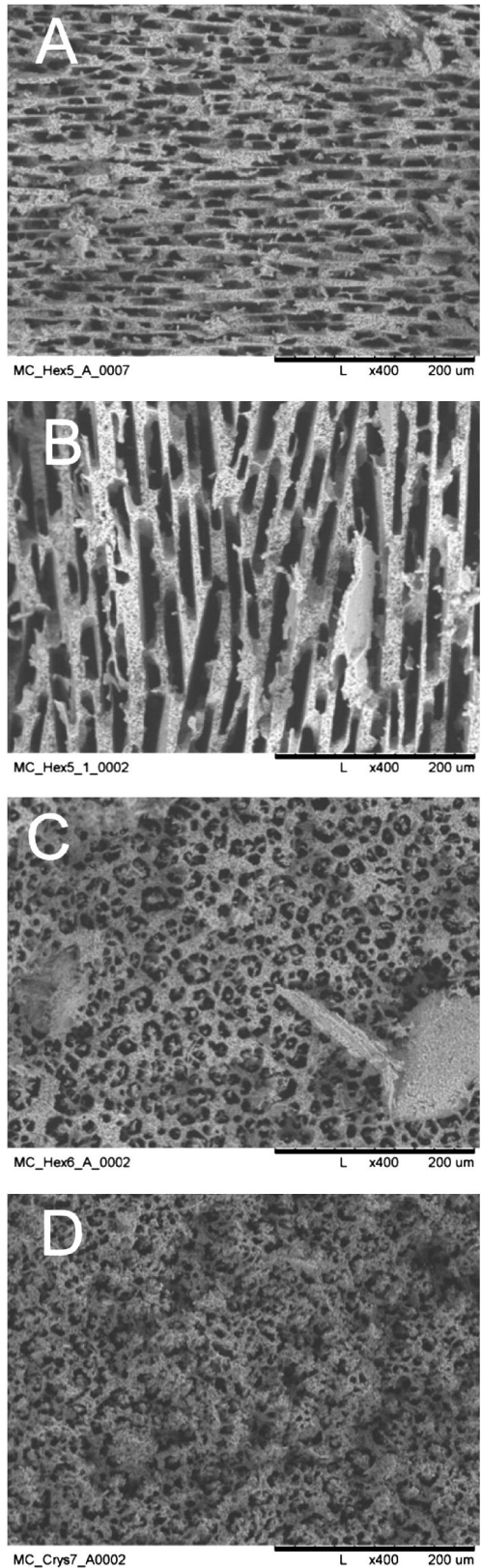

Fig. 2. SEM images of ice-templates green bodies of SiC (Hexoloy (A–C) and Crystar (D)) subjected to different parameter modifications listed in Table 1: (A) fast-freezing; (B) slow-freezing; (C) addition of ZRA–honeycomb-like structure; (D) over-loaded slurry.

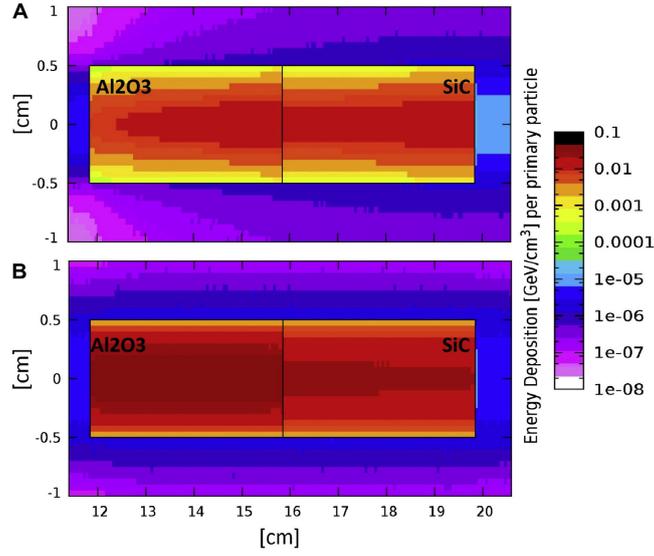

Fig. 3. Energy depostion spectra generated by FLUKA codes for (A) HiRadMat beam and (B) ISOLDE beam. In this simulation the proton beam is passing from the left to the right through the cylindrical sample partially filled with $Al_2O_3$ and SiC. The maximum of deposited energy is $2.65 \cdot 10^{-2}$ GeV/cm³/ppp (A) and $5.3 \cdot 10^{-3}$ GeV/cm³/ppp (B).

Table 2. Mechanical properties of SiC and $Al_2O_3$.

| Material | $\rho$ [g/cm3] | $E_{young}$ [GPa] | $C_p$ [J/kg.K] | $\alpha$ [1/K] $\cdot 10^{-6}$ | $\sigma_s$ [GPa] |
| --- | --- | --- | --- | --- | --- |
| SiC Crystar | 2.7 | 240 | 750 | 4.8 | 0.85 |
| SiC Hexoloy | 3.0 | 350 | 750 | 4.0 | 3.90 |
| Al2O3 | 3.8 | 370 | 880 | 8.2 | 2.60 |

Table 3. Thermo-mechanical analysis of the systems irradiated with HiRadMat and ISOLDE beam.

| Beam Material | $E_{dep}$ [J/cm3] | $E_{dep} \cdot V$ [J] | P [W] | $\Delta T$ [K] | $\sigma$ [MPa] | $\sigma^*_s$ [MPa] |
| --- | --- | --- | --- | --- | --- | --- |
| HiRadMat | | | | | | |
| SiC Hexoloy | 34 | 107 | 2.7 | 30 | 21 | 276 |
| SiC Crystar | 34 | 107 | 2.7 | 33 | 19 | 60 |
| $Al_2O_3$ | 34 | 107 | 2.7 | 20 | 31 | 184 |
| ISOLDE | | | | | | |
| SiC Hexoloy | 25 | 80 | 67 | 23 | 16 | 276 |
| SiC Crystar | 25 | 80 | 67 | 25 | 15 | 60 |
| $Al_2O_3$ | 25 | 80 | 67 | 15 | 23 | 184 |

## 3. Conclusions

It was shown that the ice-templating technique allows the design of tailor-made microstructures of pre-defined open porosity in a relatively easy and fast way. Three parameters of sample preparation can be easily tuned to independently modify different structure properties, which opens a door for testing of a large array of target material prototypes. For the first time, ice-templated microstructures are candidates for RIB production. Those prototypes have been irradiated with $10^{16}$ protons of 1.4 GeV and 440 GeV to verify if they can sustain mechanical integrity and serve as future model spallation targets. So far mechanical calculations proved that the temperature increase inside tested materials should generate thermal shock much below material's strength, therefore their microstructure should be preserved. The irradiation results will be published in a forthcoming paper.

## Acknowledgement

This research project has been supported by a Marie Curie Early Initial Training Network Fellowship of the European Community's FP7 Program under contract number (PITN-GA-2010-264330-CATHI).

## References

[1] M. Lindroos, T. Nilsson, HIE-ISOLDE: the technical options, Tech. rep., CERN-2006-013, 2006.
[2] U. Koster, ISOLDE target and ion source chemistry, Radiochim. Acta, 89 (2001), pp. 77777-77785
[3] S. Fernandes, Submicro- and nano-structured porous materials for production of high-intensity exotic radioactive ion beams, Ph.D. thesis, EPFL, Switzerland, 2010.
[4], S. Deville, Freeze-casting of porous ceramics: a review of current achievements and issues, Adv. Eng. Mater., 10 (3) (2008), pp. 155-169
[5] S. Deville, E. Saiz, A.P. Tomsia, Ice-templated porous alumina structures, Acta Mater., 55 (6) (2007), pp. 1965-1974
[6] S. Deville, C. Viazzi, J. Leloup, A. Lasalle, C. Guizard, E. Maire, J. Adrien, L. Gremillard, Ice shaping properties, similar to that of antifreeze proteins, of a zirconium acetate complex, PloS one, 6 (10) (2011), p. e26474
[7] I. Efthymiopoulos, C. Hessler, H. Gaillard, D. Grenier, M. Meddahi, P. Trilhe, A. Pardons, C. Theis, N. Charitonidis, S. Evrard, H. Vincke, M. Lazzaroni, HiRadMat: a new irradiation facility for material testing at CERN, in: Proceedings of IPAC, 2011.
[8] A. Ferrari, P.R. Sala, A. Fasso, J. Ranft, Fluka, CERN-library.

[9] L.J. Gibson, M.F. Ashby, Cellular Solids: Structure and Properties (second ed.), Cambridge University Press (1997)

[10] T. Lu, N. Fleck, The thermal shock resistance of solids, Acta mater., 46 (13) (1998), pp. 4755-4768